\documentclass[10pt,a4paper]{iopart}

\usepackage{fancybox}
\usepackage{graphicx}
\usepackage{pifont}
\usepackage{amssymb}
\usepackage{pstricks}
\usepackage{times}

\begin{document}
\jl{2}

\letter{Dynamic tunnelling ionization of H$_2^+$ in intense fields }
%\shorttitle{Dynamic tunnelling in intense fields}
\author{Liang-You Peng$^{\dag *}$, Daniel Dundas$^{\dag}$, J F McCann$^{\dag}$,
        K T Taylor$^{\dag}$ and I D Williams$^{\dag}$}

\address{$^*$ International Research Centre for Experimental Physics,\\
              Queen's University Belfast}
\address{$^{\dag}$ School of Mathematics and Physics,\\
                   Queen's University Belfast,\\
                   Belfast BT7 1NN, Northern Ireland, UK. \\
                   \vskip0.2cm
                   email : l.peng@qub.ac.uk}

\begin{abstract}
Intense-field ionization of the hydrogen molecular ion by 
linearly-polarized light is modelled by direct solution of the 
fixed-nuclei time-dependent Schr\"odinger equation and compared with 
recent experiments. Parallel transitions are calculated using 
algorithms which exploit massively parallel computers. We identify and 
calculate dynamic tunnelling ionization resonances that depend on 
laser wavelength and intensity, and molecular bond length. Results for  
$\lambda \sim 1064$ nm are consistent with static tunnelling 
ionization. At shorter wavelengths $\lambda \sim 790 $ nm large 
dynamic corrections are observed. The results agree very well with 
recent experimental measurements of the ion spectra. Our results 
reproduce the single peak resonance and provide accurate ionization 
rate estimates at high intensities. At lower intensities our results 
confirm a double peak in the ionization rate as the bond length varies.
\end{abstract}

%\section{Introduction}
\noindent
The mechanism of high-intensity ionization by infrared and optical
wavelength light is often considered  a static tunnelling process.
The simplicity of this model is hugely appealing because of the
ease of calculation. The ionization rates are effectively
independent of wavelength, and to some extent the internal
structure of the molecule can be ignored \cite{Post01}. A rough
criterion for validity of this model is given by the Keldysh
parameter,
 $\gamma_k \equiv \sqrt{|E_i|/ 2U_p}$,
 where the internal binding energy is ($|E_i|$) and the external laser-driven
kinetic energy is ($U_p$). When the conditions are such that
$\gamma_k \ll 1$,  the ionization process is dominated by static
tunnelling in which the shape of the potential strongly
(exponentially) affects the ionization rate. At certain  critical
distances between the nuclei, discovered by Codling and
co-workers~\cite{Codl93},  the ionization rate can rise sharply
producing a sequence of fast fragments ions at sharply-defined
energies. Predictions for ion yields and energies based on
classical arguments \cite{Post01} agree very well with experiments
even for large diatomic molecules such as I$_2$. The presence of
critical distances would be evident in polyatomic molecules and is
also seen in small rare-gas clusters \cite{Sied03}. The tunnelling
process is generally relatively fast compared to the vibrational
motion of the molecule, so the fixed-nuclei approximation is
reasonable. However the tunnelling time may be longer than the
optical period of the laser. Under these conditions the process is
more accurately termed a dynamic tunnelling process. In this paper
we provide evidence of just such effects for Ti:Sapphire light
$\lambda \sim 790$ nm at intensities $ I \sim 10^{14}$ W cm$^{-2}$.
Our theoretical results in this wavelength region do not agree with 
cycle-averaged static
field models, however the results do agree well with the  features
observed in experimental studies.

In a molecule with few electrons the ionization process can be studied
quantally with few approximations. For one-electron models,
static-field ionization resonances in the potential wells
~\cite{Seid95,Zuo95,Plum96,Muly96} occur at distances far from
the equilibrium internuclear separation and tend to produce
low-energy ions. Experiments have confirmed the existence of
enhanced multiphoton ionization in the hydrogen molecular ion at infrared
wavelengths \cite{Gibs97,Will00} but at intensities such that
dynamic effects of the field cannot be neglected \cite{Will03}.
 The well-established  Fourier-Floquet analysis \cite{Plum96,Plum97}
is not particularly suitable  for the study of long-wavelength excitations as
the  number of  frequency components required is very large.
Moreover, this approach supposes continuous wave conditions such
that the state decays exponentially from an isolated resonance
state with a lifetime longer than the optical cycle or natural
orbital period. Conversely, long-wavelength pulses can be
described by quasistatic fields under the conditions such that
$\gamma_k \ll 1$. However, simple tunnelling formulae  assume
exponential decay from
 a single isolated resonance connected adiabatically to the
 field-free state. This neglects nonadiabatic
 transitions within the well \cite{Muly96,Muly01} and
rescattering of the continuum electron \cite{Dund00,Dund03}. Given
these difficulties, the direct solution of the Schr\"odinger
equation has distinct advantages. It is suited for all intensities
and electronic states and  all wavelengths and pulse shapes. In
particular it is capable of describing pulse-shape effects and
nonadiabatic transitions. Thus it is highly appropriate for
realistic modelling of  experiments at infrared wavelengths such
that $\gamma_k \sim 1$.

Tunable Ti:Sapphire light $\lambda \sim 780-800$ nm interacting
with atomic hydrogen for example, achieves the pure tunnelling
regime, $\gamma_k \approx 0.1 $, only for intensities $ I >
1\times 10^{16}$W cm$^{-2}$, while for $\lambda=800$ nm  with
$I \sim 3\times  10^{14}$W cm$^{-2}$  \cite{Gibs97},  $\gamma_k
\sim 0.7$. Under these latter conditions the ionization rate is
well-defined, however a static tunnelling model is unlikely to
give a correct estimate of the resonance positions and rates; we
show that this is indeed the case. In fact, the dynamic effects
displace the critical distances, change the ionization rates, and
create electron excitation resonances. In the present work we solve the
electronic dynamics exactly by a direct solution of  the
time-dependent Schr\"odinger equation (TDSE). This does  not
include broadening due to the finite focal volume and the
corrections due to nuclear motion. 
Using atomic units, the ground state of the molecular ion is 
characterized by a bond length $R_{\rm e} = 2.0 $ and rotational
constant $B_{\rm e} = 1.36 \times 10^{-4}$. If the laser pulse duration 
is relatively short ($\sim 20$ fs compared with rotational timescale 
for this molecule ($T_{\rm rot} \sim 1/B_{\rm e} \sim 200$ fs), then 
the laser-molecule interaction can be regarded as sudden in comparison 
to the rotation of the system. Neglect of rotation effects is therefore 
reasonable. In spite of these
simplifications the results are very promising and in remarkably
good agreement with experiment for the ion energy spectrum.  The
indications are that appropriate refinements  of the model would
improve agreement, but this remains a goal for future work.

%\section{Method}

\begin{figure}[t]
\centering
\includegraphics[clip=true,width=8.0cm]{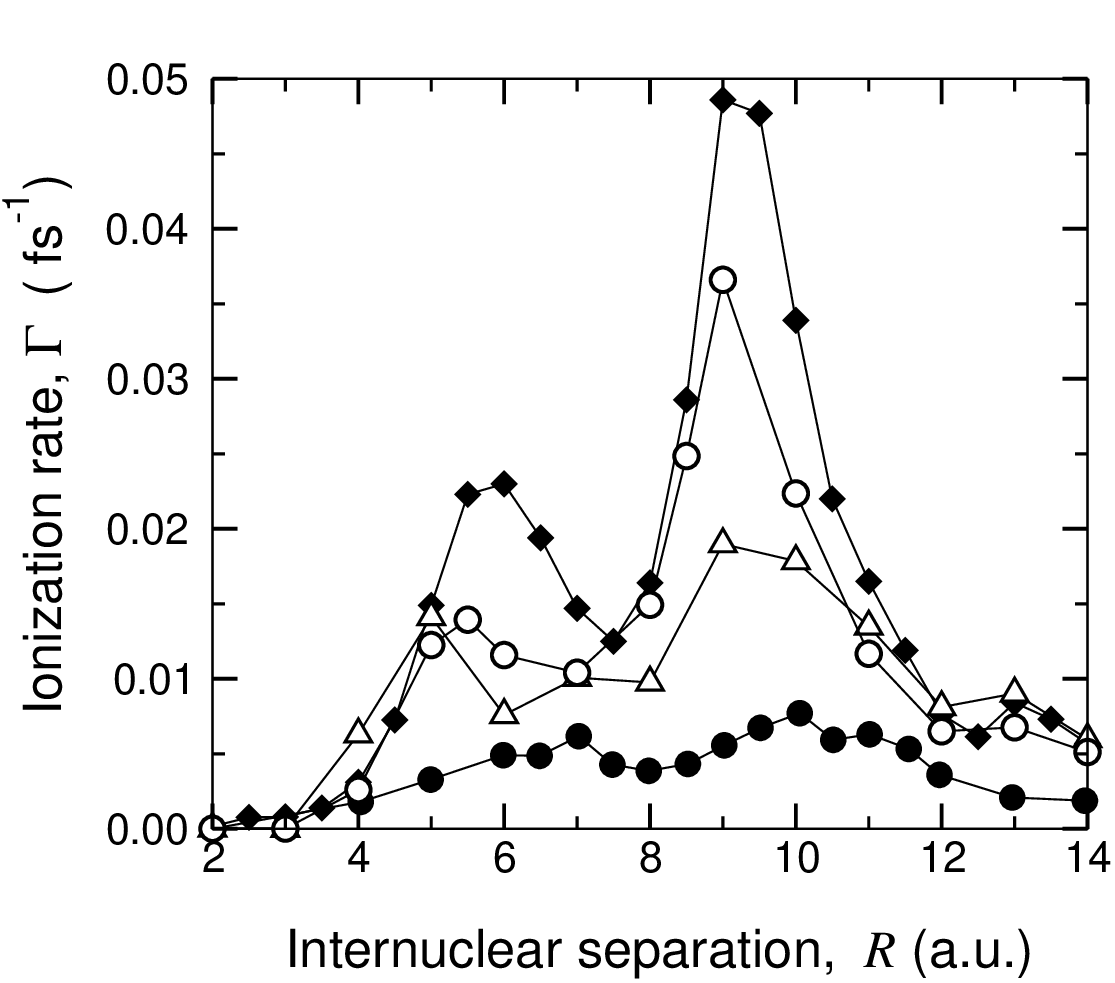}
\caption{Ionization rate, $\Gamma$,  for  dynamic  and
cycled-averaged-static  fields. The figure shows the wavelength
and bond length dependence of the ionization rate. The maximum
field strength $E_0=0.05338$ a.u. corresponds to $I
=1\times$10$^{14}$ W cm$^{-2}$. Time-independent cycle-averaged 
static field rates \cite{Plum96}, $\blacklozenge$; time-dependent rates
for $\lambda=1064$ nm (this work), $\opencircle$; 
time-dependent calculations $\lambda$=790 nm (this work), $\opentriangle$; 
time-dependent rates for $\lambda=1064$ nm \cite{Zuo95}, $\fullcircle$.} 
\label{figure1}
\end{figure}

 Making these 
approximations the TDSE, in atomic units, reads
\begin{equation}
H_{\rm e} \Psi_e( \bi{r},t) = {\rm i} {\partial \over \partial t} \Psi_e( \bi{r},t),
\end{equation}
where $H_{\rm e}$ is the electronic Hamiltonian and $\Psi_e(
\bi{r},t)$ the electronic wavefunction. Monochromatic light with
linear polarization  parallel to the internuclear axis implies a
cylindrical symmetry about the internuclear axis with the
associated good quantum number $\Lambda$. Thus the  electron
position can  be completely described by the radial, $\rho$, and
axial, $z$, coordinates with respect to an origin taken at the
midpoint between the nuclei. The TDSE reduces to a 2+1 dimensional
partial differential equation where the electronic wavefunction is
written as $\Psi_e( \rho, z, t)$ and the electronic Hamiltonian
takes the form
\begin{equation}
\fl H_{\rm e}(R;\rho,z;t) = -\frac{1}{2}
    \left( \frac{\partial^2}{\partial z^2} +\frac{\partial^2} {\partial \rho^2} +
    \frac{1}{\rho}
   \frac{\partial}{\partial \rho}\right)+ \frac{\Lambda^2}{2 \rho^2}
   + V_{\rm e}(R,\rho,z) + V_{\rm m-l}(z,t),
\end{equation}
where $R$ is the distance between the two nuclei which have charges
$Z_1$ and $Z_2$, $V_{\rm e}(R,\rho,z)$ is the electronic potential given by
\begin{equation}
     V_{\rm e}(R,\rho,z)= -{Z_1 \over \sqrt{\rho^2+(z+ \textstyle{1 \over 2} R)^2}}
     -{Z_2 \over \sqrt{\rho^2+(z-\textstyle{1 \over 2} R)^2}}.
\end{equation}
and $V_{\rm m-l}(z,t)$ the molecule-laser interaction in the
length gauge  is given by
\begin{equation}
   V_{\rm m-l}(z,t)= z E_0 f(t) \cos \omega t,
\end{equation}
where $E_0$ is the peak electric field, 
$\omega$ the angular frequency and $f(t)$ is the pulse envelope given by
\begin{equation}
   f(t)= \left\{
   \begin{array}{lc}  
      \frac{1}{2}\left[1-\cos\left(\frac{\pi t}{\tau_1}\right)\right] & 
      0\leq t\leq \tau_1\\
      1 & \tau_1\leq t \leq \tau_1+\tau_2 \\
      \frac{1}{2}\left[1-\cos\left(\frac{\pi(t-\tau_2-2\tau_1)}
      {\tau_1}\right)\right] & 
      \tau_1+\tau_2 \leq
      t\leq \tau_2+2\tau_1\\
      0 & t<0, t> \tau_2+2\tau_1
  \end{array}
  \right.,
\end{equation}
where the pulse ramp time is $\tau_1$ and the pulse duration 
$\tau_2$, with associated bandwidth $\Delta \omega= 1/\tau_2 $. In the
calculations presented in this paper $\tau_1 = 5$ cycles and $\tau_2 = 10$ 
cycles. It is convenient to
change the dependent variable to remove the first-derivative in
$\rho$ as follows~\cite{Dund02}
\begin{equation}
   \phi(\rho,z,t)= (2\pi \rho)^{1/2} \Psi_e(\rho,z,t),
\end{equation}
so that for $\Sigma$-symmetry ($\Lambda$ = 0) the time-dependent equation is
\begin{equation}
\left[ T_z +T_{\rho}
    + V_{\rm m}(\rho,z,R) + V_{\rm m-l} (z,t) \right] \phi(\rho,z,t) =
    {\rm i} \frac{\partial }{\partial t} \phi(\rho,z,t),
\end{equation}
where
\begin{equation}
    T_{\rho} \equiv -{1\over 2} \left(\frac{\partial^2} {\partial \rho^2} +
    \frac{1}{4\rho^2}\right)
    \ \ \ \ \ \ \ \ \ \ \ \ \ \ \
      T_{z} \equiv -{1\over 2} \left(\frac{\partial^2} {\partial z^2} \right),
\end{equation}
This 2+1 dimensional TDSE can be discretized on an $N_\rho \times
N_z \times N_t$ space-time grid. We label the $N_{ \rho}$ radial
grid points by, $\{\rho_1, \rho_2 , \dots \rho_{i},\dots
\rho_{N_{\rho}} \}$, while the $N_z$ axial grid points are denoted
by, $\{z_1, z_2 , \dots z_{j},\dots z_{N_z} \}$. The time
evolution progresses through the sequence of times $\{t_1, t_2 ,
\dots t_{k},\dots t_{N_t} \}$. In this case the wavefunction can
be written as the array $\phi(\rho,z,t) \rightarrow
\phi(z_i,\rho_j,t_k)$. The method of discretization of the
Hamiltonian divides the axial and radial coordinates into subspaces.
Two distinct but complementary grid methods are used for the
subspaces~\cite{Dund02}. The radial subspace is discretized on a
semi-infinite range using a small number $N_{\rho}$ of unevenly
spaced points that are the nodes of global interpolating
functions; Lagrange meshes. This leads to a small dense matrix for
the Hamiltonian in the $\rho$-subspace. On the other hand the
axial coordinate subspace is represented by a large number of
equally-spaced points, with spacing $\Delta z =0.1$ a.u.,  as
lattice points of a finite-difference scheme. The associated subspace
Hamiltonian matrix is large but sparse. Our approach is tailored
to the requirements of accuracy and computational efficiency. This
approach can easily be parallelized to make use of massively
parallel processors~\cite{Dund00,Dund02}. At the very least, the
dimensions of the cylindrical box, height $2z_{\rm max}$ radius
$\rho_{\rm max}$, must be chosen to encompass the tightly-bound
states of the system. At the same time the box should be large
enough to allow the continuum states to evolve unfettered. As the
wavefunction approaches the edge of the box boundaries, we capture
the photoelectrons by employing a masking function to absorb the
outgoing flux~\cite{Smyt98}. The $^2\Sigma^+_{\rm g}$ ground 
state is calculated via an iterative Lanzcos  calculation as 
described in~\cite{Dund02}.

The quasistatic nature of long-wavelength pulses ($\lambda \sim
1064$ nm) means that it is fair to compare the cycle-average static
field ionization rate with the time-dependent ionization rate
\cite{Muly96}. The dynamic-field (wavelength-dependent) effects
can be judged from figure \ref{figure1} in which  we choose the
wavelengths $\lambda=790\rm{nm}$ and $\lambda=1064\rm{nm}$ with
the same average intensity $I=1\times$10$^{14}$ W cm$^{-2}$
($E_0=0.05338$ a.u.). Firstly, for $\lambda=1064\rm{nm}$ the
cycle-averaged static field features  \cite{Plum96,Muly96} are
very similar to those found using our time-dependent method; with
two resonance peaks near $R \sim 5$ a.u. and $R \sim 9$ a.u.
However, there are large differences in the shape and relative
heights of the peaks. The prominent resonance near $R \sim 9$ a.u.
is the charge-resonance peak \cite{Muly01}. The inner peak ($R
\sim 5$ a.u.) is a feature of the potential barrier. The longer
wavelength $\lambda=1064$ nm does give results very similar to the
static cycle-averaged results as expected \cite{Plum96}.
For $\lambda \sim 790$ nm, there is a significant reduction 
in the heights of these peaks and some indication of peak positions moving
towards smaller internuclear separations. Calculation at 390nm demonstrate
that this trend in peak position moving to smaller values of $R$ is maintained
with the first peak found at $R=4$ a.u. for this wavelength. 
The resonance structure depends on both
bond length and wavelength. It is interesting that as the molecule
separates into its atomic fragments, the wavelength dependence
disappears and the static field result is valid.  Figure
\ref{figure1} illustrates very clearly that molecular field
ionization differs strongly from atomic field ionization for the
bond lengths, wavelengths and intensities of interest. 
Indeed the molecular ionization rates only converge to 
   within 5\% of the atomic rates at $R=20$ a.u. For instance at 
   a wavelength of 1064nm the molecular ionization rate at $R=20$ 
   is $2.85\times 10^{-3}$ fs$^{-1}$ compared with the atomic rate of 
   $2.92\times 10^{-3}$ fs$^{-1}$. These atomic rates are calculated using 
   the present code which takes $R$=0 and $Z_1 = Z_2 = 1/2$. These atomic 
   results are in agreement with other accurate time-dependent 
   results~\cite{Parker00} to within 0.5\%. Our
time-dependent results shown in figure \ref{figure1} for $\lambda
= 1064$ nm are consistent with previous static field cycle-averaged
results \cite{Plum96,Muly96}, although these results disagree with
other time-dependent results  \cite{Zuo95}. There are strong
similarities in the $R$ dependence of the rates with those calculated
previously. However our rates are up to 4 times higher than those
of \cite{Zuo95} and the resonance positions are displaced to
smaller $R$ values, so that we predict faster ion fragments with
higher yields.

\begin{figure}
\centering
\includegraphics[clip=true,width=8.0cm]{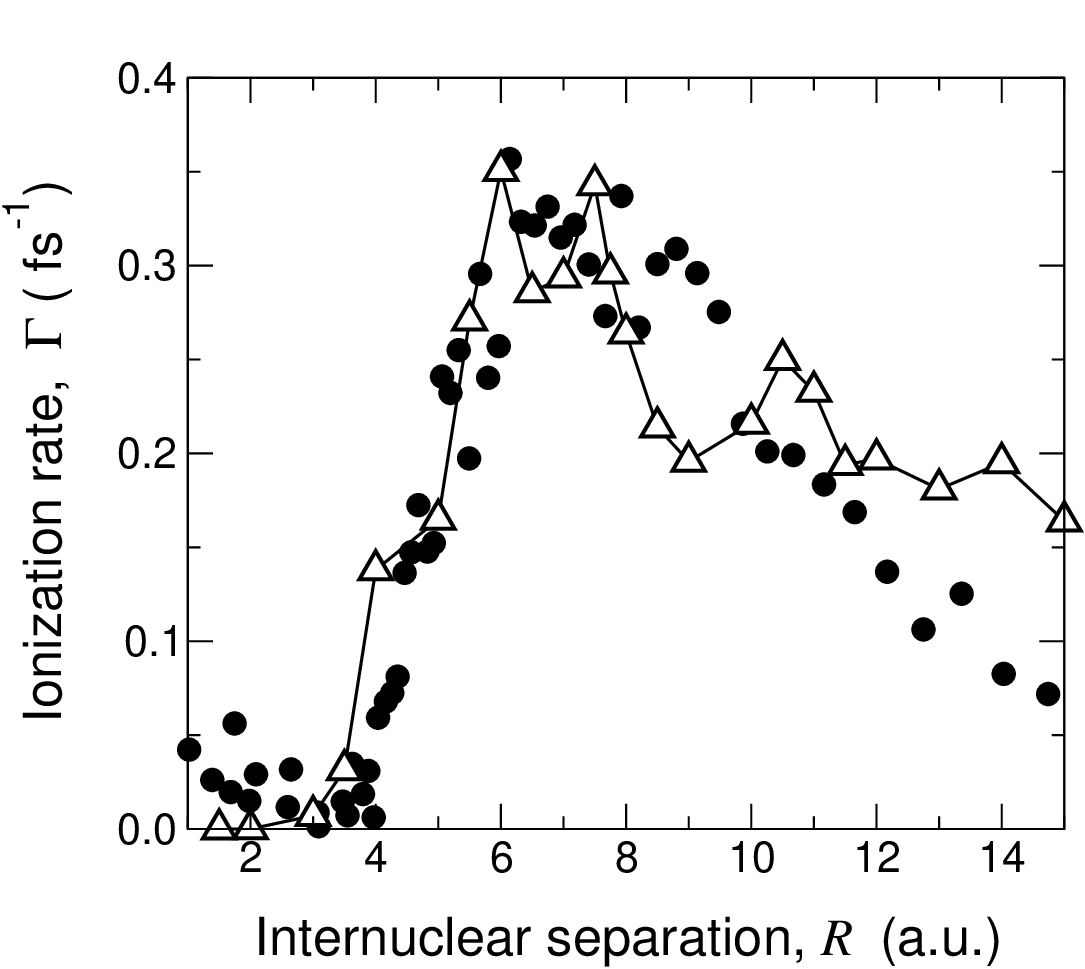}
\caption{Comparison of ionization rates with  experimental
measurements at $\lambda$=800 nm, $I =3.2\times$10$^{14}$ W
cm$^{-2}$. The experimental results ~\cite{Gibs97} ($\fullcircle$)
have been normalized to the theoretical calculations
($\opentriangle$).} \label{figure2}
\end{figure}

We apply our model to simulate experiments~\cite{Gibs97} on the
ion energy spectra from the dissociative ionization of the hydrogen 
molecular ion using $\lambda = 800$ nm at $I =3.2\times$10$^{14}$ W
cm$^{-2}$; the results are presented in figure \ref{figure2}. In
the experiment an H$_2$ gas was ionized to form the H$_2^+$ ion
target. The molecular ions subsequently dissociate and ionize in
the laser field, with the proton fragments extracted and energy
analyzed. By relating the kinetic energies of the ions to the
Coulomb explosion curve, ionization rates could be deduced for the
range of molecular bond lengths. The experimental conditions were
such that saturation of the ionization channel was avoided,
permitting ion yields from larger $R$ values to be estimated and
eliminating ion yields from the larger focal volume. Our
approximation in assuming a very localized high intensity region,
smaller than the diffraction limit, is well justified. The sensitivity 
of the ionization rate to changes in intensity and wavelength was 
noted in the results presented in figure \ref{figure1}.  
Consider the changes in the ionization rates in going from the data 
presented in figure~\ref{figure1} for $\lambda=790$ nm to that for 
$\lambda=800$ nm at an intensity 3.2 times higher, in 
figure~\ref{figure2}. The ionization rates in figure~\ref{figure2} are 
roughly 15 times higher which is consistent with an
exponential increase giving rise to more easily measurable ion
yields. However the double peak structure of figure \ref{figure1}
is now dominated by a single broad maximum near $R \sim 7$. In
comparing with experiment we have in figure~\ref{figure2} normalized 
the laboratory data to our results. We see that the shape of the 
theory and experimental curves are in remarkable agreement, in spite 
of the assumptions made. The single broad peak is reproduced rather 
well, although some additional structure present in the simulations 
is not resolved by experiment. For $R>8$ the theory and experiment 
are in disagreement, and the  theoretical estimate of ion yield from 
$R \sim 9$ is much lower than the experimental results. This might be
attributed in part to the variation in focal volume intensity. At
the edges of the focal spot the intensities decrease but the
interaction volume is larger \cite{Post01}. Moreover at lower
intensities the field ionization rates move to larger $R$
\cite{Plum96}. So one would expect that an inclusion of focal
volume variation would broaden the peak to larger values of $R$
and partially compensate for this shortfall. The second feature is
that the theoretical results predict high ionization far in excess of
that found experimentally at large $R$. The $R \rightarrow \infty$ atomic 
limit of the theory results is very accurately known and consistent with
the theoretical results in figure \ref{figure2}. The theoretical data for
ionization rates can be considered as accurate. However the lower
ion yield observed can be explained by the fact that during 
molecular dissociation the molecular ion can ionize at smaller $R$
values. If the ionization rates are large at small $R$, then few if any
molecules can survive to be ionized at large bond lengths. A rough 
estimate of the survival probability $P(R)$ of the molecular ion
can be found from the classical dynamics of the ionized molecular ion.
The depletion rate is given by $dP/dR \approx -
(\Gamma /v) P$ where $v(R)$ is the classical relative velocity of
the protons such that $m_pv^2/4 + Z_1Z_2/R = Z_1Z_2/R_e$ and 
$m_p$ is the proton mass. For the case $\lambda =800$ nm and 
$I = 3.2\times10^{14}$ Wcm$^{-2}$ a rough estimate based on this model gives $P\approx0.2$
at $R = 14$. This is only an indication 
of the effect but it is consistent with the findings of Dundas~\cite{Dund03}.

\begin{figure}
\centering
\includegraphics[clip=true,width=8.0cm]{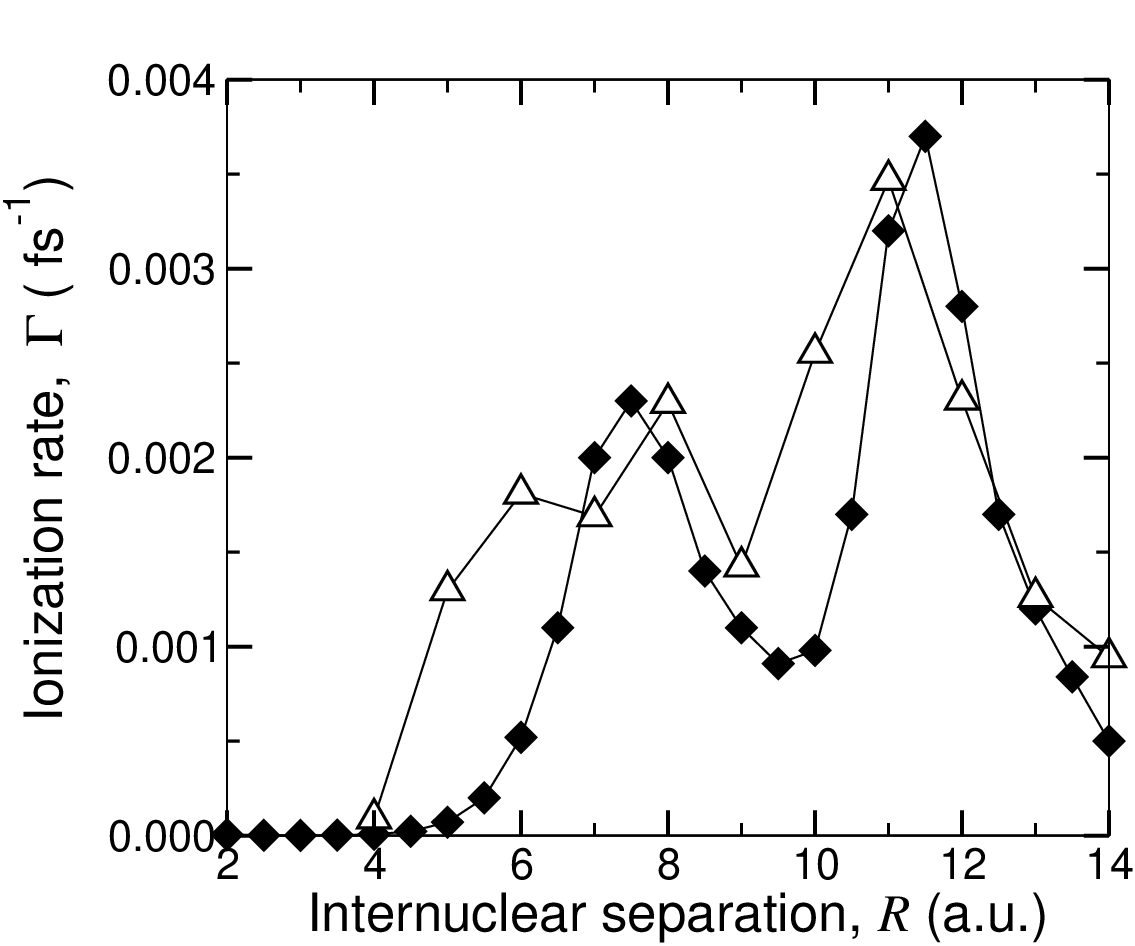}
\caption{Comparison of ionization rates from the present
time-dependent calculations ($\opentriangle$) at $\lambda$=790 nm
and $I =0.6\times10^{14} $W cm$^{-2}$ ($E_0=0.041$ a.u.) with
static field cycle-average rates \cite{Plum96} ($\blacklozenge$) at a
comparable intensity of $I =0.56\times10^{14} $W cm$^{-2}$
$E_0=0.040$ a.u.} \label{figure3}
\end{figure}

Very recently data have become available from experiments in
Garching on H$_2^+$ for  $\lambda=790\rm{nm}$ at intensities just
above the Coulomb explosion threshold, namely $I=0.6\times
10^{14}$ W cm$^{-2}$~\cite{Pavi03}. In these results the first
observations of vibrationally resolved structure has been
obtained. From the ion momentum distribution, it was suggested that
a critical distance around $R=12$ could explain the results.
Existing static field cycle-average rates at a comparable
intensity $I =0.56\times10^{14} $W cm$^{-2}$, $E_0=0.04$ a.u.
\cite{Plum96} are shown in figure~\ref{figure3}. These results
confirm that the ionization rates are extremely small, the
reduction in intensity by a factor of 5 leading to ion yields
roughly one hundred times smaller. The double peak structure
emerges in our calculations, and since the rates are now reduced, 
the bulk of the
molecules will reach the outer resonance position. In figure
\ref{figure3} the time-dependent calculations for $\lambda$=790 nm
and $I =0.6\times10^{14} $W cm$^{-2}$ are in fairly good agreement
with the static field results. However we note (figure
\ref{figure3}) the inner peak $R \sim 6-8$ is broad and high and
ought to produce ion yields comparable to the sharp outer peak
near  $R=11$. The observation of quantal vibrational structure in
the ion spectrum means that a full quantum treatment of nuclear
dynamics is required to analyze these new experiments in full.

%\section{Conclusions}

We have solved the full-dimensional TDSE for the electron dynamics of
H$_2^+$ in linearly polarized laser fields, assuming that the nuclei
are fixed in space. The method employed is highly accurate and
can be efficiently implemented on parallel processing computers.
Ionization rates can be calculated for all nuclear
separations and for wavelengths
from the infrared to x-ray, for a range of laser pulses.
Comparison of our results with other theoretical calculations
and recent experimental measurements
show very good agreement. We have been able to identify and calculate
dynamic tunnelling resonances for
$\lambda =790$ nm  and $\lambda =800$ nm
and obtain accurate estimates of the ionization rates and
large dynamic tunnelling corrections are observed.
A major simplification in the model is the fixed-nuclei
assumption. However, our results for  $\lambda =800$ nm and
$I=3.2\times 10^{14}$ W cm$^{-2}$
reproduce the measured dependence of ionization rate on bond length.
At shorter wavelength and
lower intensities,  $\lambda =790$ nm and
$I=0.6\times 10^{14}$ W cm$^{-2}$, our results indicate a double peak
structure in the ionization rate as the bond length varies. The outer
ionization resonance agrees with experimental measurements~\cite{Pavi03}.

To model the experiments more realistically several extensions to
the current approach are required. Firstly, the energy and angular 
momentum exchanges between the nuclei and electrons will occur during 
process. Dundas \cite{Dund03} has
combined the full electronic dynamics with a quantal vibrational
motion for intense field dissociative ionization and found that
the dynamic tunnelling resonances dominate strongly over pure
dissociation at high intensities. A classical model of nuclear
motion will not be sufficient as the wavepacket will disperse
during the process and indeed experiments are now able to resolve
the vibrational structure in the ion yield \cite{Pavi03}. The
quantal motion is essential to obtain an ion spectrum distribution
rather than one-to-one mapping of ion energies to specific bond
lengths. Secondly, within the laser focal spot, there is a spatial
variation of intensity which has to be taken into account above the 
saturation intensity. Thirdly, while present calculations only 
consider parallel electronic transitions, we must consider results 
averaged over molecular orientation. 
Previous work by Plummer and McCann~\cite{Plum97} found that 
DC ionization rates decrease sharply as the angle of orientation of 
the the molecular axis with the field increases. The orientation 
dependence of dynamic tunnelling ionization has yet to be established.
These refinements are likely to be more important in the very high
intensity regime $ I \sim 10^{15}$ W cm$^{-2}$ rather than the
regime $ I \sim 10^{14}$ W cm$^{-2}$.  We intend to undertake
refinements of our model to simulate these effects and produce
accurate estimates of ion yields and ion energy spectra.

LYP acknowledges the award of a PhD research studentship from
the International Research Centre for Experimental Physics, Queen's
University Belfast. DD acknowledges the award of an EPSRC Postdoctoral
Fellowship in Theoretical Physics. This work has also been supported 
by a grant of computer resources at the Computer Services for Academic 
Research, University of Manchester, provided by EPSRC to the UK Multiphoton,
Electron Collisions and BEC HPC Consortium.

%\begin{references}
\section*{References}
%\begin{harvard}

%\end{harvard}
%\end{section}

\end{document}